%% file: Gflux.tex
\documentstyle[12pt]{article}
\input lmacros

\def\Im{\mop{Im}}
\def\comment#1{  \begin{raggedright}{\tt [#1]}\end{raggedright}}
\def\tend#1{{\hat{#1}}}
\def\fourd#1{{#1}}
\def\sixd#1{{\tilde{#1}}}
\def\flat#1{{\underline{#1}}}
\def\conj#1{{\overline{#1}}}
\def\e{{\rm e}}   
\begin{document}
\baselineskip=15.5pt
\pagestyle{plain}
\setcounter{page}{1}
\renewcommand{\thefootnote}{\fnsymbol{footnote}}

\begin{titlepage}

\begin{flushright}
PUPT-1957 \\
hep-th/0010010
\end{flushright}
\vfil

\begin{center}
{\Large Supersymmetry and F-theory realization}
\vskip0.5cm
{\Large of the deformed conifold with three-form flux}
\end{center}

\vfil
\begin{center}
{\large Steven S. Gubser}
\end{center}

$$\seqalign{\span\TL & \span\TT}{
& Joseph Henry Laboratories, Princeton University, Princeton,
NJ 08544
}$$
\vfil

\begin{center}
{\large Abstract}
\end{center}

\noindent 
 It is shown that the deformed conifold solution with three-form flux,
found by Klebanov and Strassler, is supersymmetric, and that it admits
a simple F-theory description in terms of a direct product of the
deformed conifold and a torus.  Some general remarks on Ramond-Ramond
backgrounds and warped compactifications are included.

\vfil
\begin{flushleft}
October 2000
\end{flushleft}
\end{titlepage}
\newpage

\section{Introduction}
\label{Intro}

It is believed on general grounds \cite{PolCave} that a string theory
capable of describing the low-energy limit of QCD should be defined on
a warped background in some dimension larger than four:
  \eqn{WarpedGeometry}{
   d\tend{s}^2 = e^{2A(y)} \fourd\eta_{\mu\nu} dx^\mu dx^\nu - 
    e^{-2A(y)} \sixd{g}_{IJ} dy^I dy^J \,,
  }
 where we use mostly minus signature for the $3+1$-dimensional
Minkowski metric $\fourd\eta_{\mu\nu} dx^\mu dx^\nu$ and for the
higher dimensional metric $d\tend{s}^2$, and all plus signature for
the extra-dimensional metric $\sixd{g}_{IJ} dy^I dy^J$.  The choice of
warp factor $e^{-2A(y)}$ on the extra dimensions will prove convenient
later, although it could have been absorbed into $\sixd{g}_{IJ}$.  In
seeking for supersymmetric solutions of this form in critical string
theory (so that $\sixd{g}_{IJ} dy^I dy^J$ represents a six-dimensional
metric), Ramond-Ramond fields seem a necessary ingredient.

There does not seem to be a clean general argument to the effect that
type II string backgrounds dual to gauge theories must involve
Ramond-Ramond fields.  There is one, however, if we assume minimal
supersymmetry in $3+1$ dimensions.  Supersymmetric solutions with only
the bosonic fields in the NS-NS sector turned on (namely, the metric,
the dilaton, and the two-form potential) have been shown to have an
unwarped string metric \cite{andy}.  This amounts to the statement
that the worldsheet CFT factorizes into a $3+1$-dimensional part and
an additional piece from the six extra dimensions.  Zig-zag symmetry
cannot be implemented unless there is a warp factor which either
diverges or vanishes at some point \cite{PolCave}.  Thus we must
consider Ramond-Ramond backgrounds if a string dual to
super-Yang-Mills theory is desired.  

An interesting solution exhibiting some of the features of confinement
(namely the area law for Wilson loops, screening for appropriately
defined magnetic flux, and a mass gap) was recently exhibited in
\cite{ks}.  A somewhat similar solution appeared in \cite{mn} as a
lift of a solution to seven-dimensional supergravity
\cite{chamseddine}.  Supersymmetry was not demonstrated in \cite{ks}.
The more abstract and general treatment via superpotentials that
appeared in \cite{VafaTop} makes it clear that a supersymmetric
solution exists.  The aim of this note is to demonstrate that the
solution of \cite{ks} is indeed supersymmetric, that the simple first
order equations appearing in \cite{ks} are precisely the conditions
for supersymmetry, and that the geometry is an explicit example of an
F-theory compactification on the product of a non-compact Calabi-Yau
three-fold and a torus.  These results were independently derived in
\cite{PolchinskiGrana}.

\section{Supersymmetry in type IIB supergravity}
\label{SUSYIIB}

The tools for studying bosonic backgrounds of ten-dimensional type IIB
supergravity are the bosonic equations of motion and the fermionic
supersymmetry variations.  They are, in a unitary gauge for the
$SL(2,{\bf R})/U(1)$ coset,
  \eqn{IIBEOMS}{\eqalign{
   \tend{D}^M P_M &= {\kappa^2 \over 24} G_{MNP} G^{MNP}  \cr
   \tend{D}^P G_{MNP} &= P^P G^*_{MNP} - {2i \over 3} \kappa
     F_{MNPQR} G^{PQR}  \cr
   \tend{R}_{MP} &= P_M P^*_P + P^*_M P_P + {\kappa^2 \over 6}
     F_{Q_1 \cdots Q_4 M} F^{Q_1 \cdots Q_4}{}_P  \cr
   &\quad {} + {\kappa^2 \over 8} \left( G_M{}^{QR} G^*_{PQR} + 
    G^*_M{}^{QR} G_{PQR} - {1 \over 6} \tend{g}_{MP} G^{QRS} G^*_{QRS}
     \right)  \cr
   F_{(5)} &= \tend{*} F_{(5)}
  }}
  \eqn{IIBSUSY}{\eqalign{
   \delta\lambda &= {i \over \kappa} P_M \tend\gamma^M \epsilon^* - 
    {i \over 24} G_{MNP} \tend\gamma^{MNP} \epsilon  \cr
   \delta\psi_M &= {1 \over \kappa} \hat{D}_M \epsilon + 
     {i \over 480} F_{P_1 \cdots P_5} \tend\gamma^{P_1 \cdots P_5}
     \tend\gamma_M \epsilon + {1 \over 96} (\tend\gamma_M{}^{NPQ} G_{NPQ} - 
     9 \tend\gamma^{NP} G_{MNP}) \epsilon^*
  }}
 where
  \eqn{IIBdefs}{\eqalign{
   &F_{(5)} = dA_{(4)} - {\kappa \over 8} \Im A_{(2)} \wedge F^*_{(3)} \qquad
   F_{(3)} = dA_{(2)}  \cr
   &G_{(3)} = {F_{(3)} - B F^*_{(3)} \over \sqrt{1-|B|^2}}  \qquad
   P_M = {\partial_M B \over 1-|B|^2}  \cr
   &\tend\gamma_{11} \lambda = \lambda \qquad
    \tend\gamma_{11} \psi_M = -\psi_M \qquad
    \tend\gamma_{11} \epsilon = -\epsilon \,.
  }}
 Except for some typographic alterations, the conventions used in this
note are those of \cite{SchwarzIIB}; in particular, the metric
signature is mostly minus.  All explanation of notation is relegated
to the Appendix.

In \cite{Kehagias}, solutions to the supersymmetry transformations
laws were considered where the three-form was set to zero but the
scalars could vary.  Here we wish to do the opposite and consider
constant scalars.  This means $G_{MNP} G^{MNP} = 0$, which is certainly
satisfied if $\sixd{*} G_{(3)} = i G_{(3)}$.  This latter
``self-dual'' ansatz is the one we wish to focus on.  In purely
ten-dimensional terms, $\tend{*} G_{(3)} = i e^{4A} \vol_4 \wedge
G_{(3)}$.  Without loss of generality we can choose $B=0$ and then
apply a global $SL(2,{\bf R})$ transformation to restore arbitrary $B$.

A compact rewriting of the gravitino variation in \IIBSUSY\ is as follows:
  \eqn{DelGravitino}{\eqalign{
   \delta\lambda &= {i \over \kappa} \slashed{P}_{(1)} \epsilon^* - 
    {i \over 4} \slashed{G}_{(3)} \epsilon  \cr
   \delta\psi_M &= {1 \over \kappa} \tend{D}_M \epsilon + 
    {i \over 4} \slashed{F}_{(5)} \tend\gamma_M \epsilon -
    {1 \over 16} (2 \slashed{G}_{(3)} \tend\gamma_M + 
     \tend\gamma_M \slashed{G}_{(3)}) \epsilon^* \,.
  }}
 The first equation is easy to solve because $\slashed{P}_{(1)} = 0$
and $\slashed{G}_{(3)} {1 - i \sixd\gamma_7 \over 2} =
\slashed{G}_{(3)}$ (this last equation is a re-writing of $\sixd{*}
G_{(3)} = i G_{(3)}$).  The result is that any spinor with
$\sixd\gamma_7 \epsilon = -i \epsilon$ will satisfy the dilatino
variation equation.  The largest possible holonomy group for the
internal manifold is $SO(6) \approx SU(4)$, and from $\sixd\gamma_7
\epsilon = -i \epsilon$ we learn that $\epsilon$ falls in the ${\bf
4}$ of $SU(4)$ rather than the $\bar{\bf 4}$.  Note that complex
conjugation of a spinor reverses the eigenvalue of $\sixd\gamma_7$, as
does multiplication by any $\tend{\gamma}_I$.

The next equation to look at is the gravitino variation in the extra
dimensions, which simplifies to
  \eqn{GravitinoFirst}{
   \delta\psi_I = {1 \over \kappa} \tend{D}_I \epsilon + 
    {i \over 4} \slashed{F}_{(5)} \tend\gamma_I \epsilon -
    {1 \over 16} \tend\gamma_I \slashed{G}_{(3)} \epsilon^* \,.
  }
 The form of $F_{(5)}$ is restricted by $3+1$-dimensional Poincar\'e
invariance and self-duality: it is
  \eqn{FAssume}{
   F_{(5)} = h_{(5)} + \tend{*} h_{(5)}
  }
 where $h_{(5)} = -\sixd{*} h_{(1)}$ is a five-form on the extra
dimensions and $\tend{*} h_{(5)} = e^{8A} \vol_4 \wedge h_{(1)}$.  To
satisfy the Bianchi identity for $F_{(5)}$ we must have
  \eqn{BianchiF}{\eqalign{
   d h_{(5)} &= -{\kappa \over 8} \Im F_{(3)} \wedge F_{(3)}^*  \cr
   d \tend{*} h_{(5)} &= 0 \,.
  }}
 Multiplying \GravitinoFirst\ by $1 \pm i \sixd\gamma_7$ one sees that
the first two terms must cancel against one another, and the last term
must vanish on its own.  Thus we can separate the conditions on
$G_{(3)}$,
  \eqn{GthreeEqns}{
   \sixd{*} G_{(3)} = i G_{(3)} \,, \qquad
   \slashed{G}_{(3)} \epsilon = \slashed{G}_{(3)} \epsilon^* = 0 \,,
  }
 from the rest of the conditions for supersymmetry, 
  \eqn{PsiSimpler}{
   \delta\psi_M = \left[ {1 \over \kappa} 
    \left( \partial_M + {1 \over 4} \tend\omega_{M\flat{NP}} 
     \tend\gamma^\flat{NP}\right) - e^{4A} {i \fourd\gamma_5 \over 2} 
     h_J \tend\gamma^J \tend\gamma_M \right] \epsilon = 0 \,,
  }
 which, in terms of a rescaled spinor $\tilde\epsilon = e^{-A/2}
\epsilon$ whose eigenvalue under $\fourd\gamma_5$ is $-i$, read
  \eqn{PsiSimplest}{\eqalign{
   \delta\psi_\mu &= e^{A/2} \tend\gamma_\mu{}^J 
    \left[ {1 \over 2\kappa} \partial_J A - 
     {e^{4A} \over 2} h_J \right] 
      \tilde\epsilon = 0  \cr
   \delta\psi_I &= e^{A/2} \left[ {1 \over \kappa} \sixd{D}_I + 
    \left( {1 \over 2\kappa} \partial_I A - 
     {e^{4A} \over 2} h_I \right) - 
    \tend\gamma_I{}^J \left( {1 \over 2\kappa} \partial_J A -
     {e^{4A} \over 2} h_J \right) \right] 
      \tilde\epsilon = 0 \,.
  }}
 Equivalent to \PsiSimplest\ are the conditions
  \eqn{PsiSolve}{
   h_{(1)} = -{1 \over 4\kappa} d e^{-4A} \,, \qquad
   \sixd{D}_I \zeta = 0 \,.
  }
 Thus we learn that $\sixd{g}_{IJ}$ is a metric of (at most) $SU(3)$
holonomy, which is to say a Calabi-Yau metric.  If the holonomy is
exactly $SU(3)$, then the only possible choice for $\tilde\epsilon$ is
the $SU(3)$ singlet with eigenvalue $-i$ under $\sixd\gamma_7$,
multiplied by an arbitrary spinor in the ${\bf 2}$ of $SO(3,1)$.  Thus
four supercharges are preserved: ${\cal N}=1$ in four dimensions, as
implicitly claimed in \cite{ks}.  Indeed, (81) and (82) of \cite{ks}
are precisely the supersymmetry conditions $\sixd{*} G_{(3)} = i
G_{(3)}$ and $h_{(1)} = -{1 \over 4\kappa} d e^{-4A}$.  If the
holonomy is $SU(2)$, then eight supercharges are preserved.

It remains only to check the Bianchi identities, \BianchiF.  Using
\PsiSolve\ one obtains $\tend{*} h_{(5)} = {1 \over 4\kappa} \vol_4
\wedge d e^{4A}$, which is obviously closed.  Thus the first equation
in \BianchiF\ gives us our only constraint: in two equivalent forms,
  \eqn{BianchiConstraint}{\eqalign{
   d \sixd{*} h_{(1)} &= {\kappa \over 8} \Im F_{(3)} \wedge F_{(3)}^*  \cr
   \sixd\square e^{-4A} &= {\kappa^2 \over 2} \sixd{*} 
    \Im F_{(3)} \wedge F_{(3)}^* = -{\kappa^2 \over 12}
    \sixd{g}^{I_1 J_1} \sixd{g}^{I_2 J_2} \sixd{g}^{I_3 J_3}
     F_{I_1 I_2 I_3} F^*_{J_1 J_2 J_3} \,.
  }}
 To obtain the second form we have used \PsiSolve\ and $\sixd\square =
\sixd{D}_I \sixd{D}^I = -\sixd{*} d \sixd{*} d$ acting on scalars.
Keeping track of all the signs is somewhat difficult, but there is a
consistency check: the second form of \BianchiConstraint\ is exactly
the trace of the Einstein equations.  Unsurprisingly, it is impossible
to satisfy \BianchiConstraint\ on a compact manifold unless $F_{(3)} =
0$: the right hand side is negative and the left hand side is a total
derivative.

We have stated in complete detail the conditions for supersymmetry in
equations \GthreeEqns, \PsiSolve, and \BianchiConstraint.  Because the
the six extra dimensions form a complex manifold, the conditions
\GthreeEqns\ can be simplified.  The equation $\sixd{*} G_{(3)} = i
G_{(3)}$ is satisfied precisely if $G_{(3)}$ is a sum of a $(2,1)$
form and a $(0,3)$ form (both closed of course).  The $SU(3)$ singlet
spinor can conveniently be defined as the Fock space ground state
annihilated by $\sixd\gamma^{\flat{p}}$, where the holomorphic vector
index $p$ runs from $1$ to $3$.  If $G_{(3)}$ contains a $(0,3)$
component, then $\slashed{G}_{(3)}$ fails to annihilate the Fock
vacuum, since it contains a term proportional to the product
$\sixd{\gamma}^{\flat{\bar{1}}} \sixd{\gamma}^{\flat{\bar{2}}}
\sixd{\gamma}^{\flat{\bar{3}}}$ of all three creation operators.  So
$G_{(3)}$ must be a $(2,1)$ form.

Generally speaking, it is not trivial to find a closed $(2,1)$ form
for which \BianchiConstraint\ can be solved to obtain $e^{-4A}$ which
is everywhere non-singular and vanishes at infinity.  This is exactly
what the authors of \cite{ks} did for the case of the deformed
conifold.  It would be quite interesting to inquire in what generality
non-singular solutions to \BianchiConstraint\ exist.  Whenever one is
found, it is always possible to add an arbitrary function to $e^{-4A}$
which is harmonic except for delta function sources and 
vanishes at infinity.  This corresponds to adding D3-branes at
arbitrary locations, and doesn't break any additional supersymmetry.

\section{Relation to F-theory}
\label{Ftheory}

A number of authors \cite{vwGukov,Sethi,Mayr,VerlindeChan} have
considered warped F-theory compactifications related to M-theory on
Calabi-Yau four-folds with G-flux \cite{Becker}.\footnote{Type~IIB
vacua with three-form flux have also been studied in
\cite{VafaTaylor}.}  It is straightforward to see how the solutions
described in the previous section fit into this rubric.  The general
prescription is that F-theory on an elliptically fibered $\hbox{CY}_4$
is equivalent to type IIB string theory on the base of the fibration,
where the IIB coupling is identified with the modular parameter of the
torus.  In our case, this coupling is constant, so we must be
considering F-theory on a product of $T^2$ and the Calabi-Yau
three-fold with metric $\sixd{g}_{IJ} dy^I dy^J$.  We have restricted
attention to the case where the coupling is $B=0$, or equivalently
$\tau=i$: this corresponds to a square $T^2$.  Other complex
structures can be obtained through a global $SL(2,{\bf R})$ rotation.

In the construction of M-theory compactifications on eight-manifolds
with G-flux \cite{Becker}, it was shown that $G_{(4)}$ had to be a
$(2,2)$-form.  In translating the related F-theory compactification
into type~IIB language, the following formula is standard (see for
instance \cite{vwGukov,Sethi,VerlindeChan}):
  \eqn{StandardG}{
   G_{(4)} = {\pi \over i \Im \tau} (H \wedge d\bar{z} - 
    \bar{H} \wedge dz) \,,
  }
 where $H = H^{R} - \tau H^{NS}$ and $z$ is the holomorphic coordinate
on the $T^2$.  Since $\tau = i$ for us, we may identify $H =
G_{(3)} = F_{(3)}$.  From \StandardG\ we learn that for $G_{(4)}$ to
be a $(2,2)$ form, $G_{(3)}$ must be a $(2,1)$ form---as concluded
earlier from a direct analysis in the type IIB language.  Furthermore, the
equation for the warp factor, \BianchiConstraint, descends from an
analogous formula ((2.57) of \cite{Becker}), which for the warped
M-theory geometry
  \eqn{MGeom}{
   ds^2 = e^{-\phi(y)} \eta_{\mu\nu} dx^\mu dx^\nu + 
    e^{\phi(y)/2} g_{IJ} dy^I dy^J
  }
 reads
  \eqn{MForm}{
   \square e^{3\phi/2} = * \left[ 4\pi^2 X_8(R) - 
    {1 \over 2} G \wedge G - 4\pi^2 \sum_{i=1}^n \delta^8(y-y_i)
    \right]
  }
 where $\square$ and $*$ are defined with reference to the Calabi-Yau
metric $g_{IJ}$ on the eightfold, as is
  \eqn{XEight}{
   X_8 = {1 \over (2\pi)^4} \left[ -{1 \over 768} (\tr R^2)^2 + 
    {1 \over 192} \tr R^4 \right] \,.
  }
 In translating \MForm\ through F-theory to type~IIB, the last term
changes from source terms for M2-branes to source terms for D3-branes.
The first term goes away when $\tau$ is constant because $X_8$
vanishes for the product of a Calabi-Yau three-fold and $T^2$.  Thus
F-theory considerations do {\it not} lift the topological obstruction
to solving \BianchiConstraint\ on a compact manifold, unless we lift
the assumption that the complex coupling is constant.  In a more
general F-theory compactification on an elliptically fibered
$\hbox{CY}_4$ with nonzero Euler number (and, necessarily, nontrivial
fibration of the $T^2$), the global constraint that arises from the
generalization of \BianchiConstraint\ is
  \eqn{FTheoryChi}{
   {\chi \over 24} = n_{D3} + \int \Im G_{(3)} \wedge G_{(3)}^*
  }
 in appropriate units.  (The Euler number $\chi$ vanishes when the
eight-manifold is a Calabi-Yau times $T^2$, and then there is a
problem because both terms on the right hand side are positive).  Such
compactifications have been studied \cite{VerlindeChan,Mayr} as
possible realizations of the Randall-Sundrum scenario
\cite{rsOne,rsTwo}.

Minimally supersymmetric compactifications of F-theory lie at the
center of a fascinating locus of ideas.  As warped geometries, they
offer the hope of obtaining a hierarchy of scales from
geometry.\footnote{The generality of this idea as an extension of
\cite{rsOne} has been emphasized to me by H.~Verlinde.}  It was
observed in \cite{Sethi} that the conditions for supersymmetry are
difficult to satisfy and admit few moduli compared to usual
Calabi-Yau compactifications.  The discovery \cite{ks} that
a deformation of the conifold was necessary in order to have G-flux
resolve the naked singularity found in \cite{kt} can be viewed as an
example of this moduli fixing, since the size of the $S^3$ at the tip
of the deformed conifold is determined by the G-flux.  Warped
compactifications of F-theory have even been speculated to offer a
solution \cite{Vafa} to the cosmological constant problem along the
lines of \cite{WittenThreeD}.  The eventual hope is to find an
isolated compactification, with strong warping to account for the
hierarchy between the gravitational and electroweak scales, and broken
supersymmetry without a large cosmological constant.

The properties of F-theory compactifications, and in particular the
claims of \cite{ks}, suggest that there is a big chunk missing from
our understanding of type~II string theory.  It was suggested in
\cite{ks} that the deformed conifold with three-form represented a
duality cascade of $SU(M) \times SU(N)$ gauge theories.  The logic
originated with a study of related geometries in AdS/CFT
\cite{juanAdS,gkPol,witHolOne}: first there was the idea
\cite{gkBaryon} that D5-branes wrapped on a two-cycle of $T^{11}$
represented domain walls between a $SU(N) \times SU(N)$ gauge theory
and a $SU(N) \times SU(N+1)$ gauge theory; then there were
supergravity treatments of the geometry that would arise from a small
number of such ``fractional branes'' (in the sense of \cite{Douglas})
added to many D3-branes \cite{kn}; then there came the extension
\cite{kt} to the case of only fractional branes, which finally led
\cite{ks} to a wholly non-singular geometry without any D-branes at
all, just three-form flux.  The origins of the construction lead us to
believe that the final geometry has a large hidden gauge symmetry,
which originated in the Chan-Paton factors of the open strings
attached to the fractional branes that are smoothed away in the end.
And it is natural to believe that this hidden gauge symmetry persists
in a compactification of F-theory which locally looks like the
deformed conifold solution of \cite{ks}.  Intuitively, the open
strings are there, just confined---or, perhaps more appropriately,
``dualized'' into a smooth closed string background.  AdS/CFT thus
seems to lead us toward a vastly more general open-closed string
duality, applicable (one would hope) to compact as well as non-compact
geometries.  Studies of tachyon condensation \cite{Piljin,Sen} seem to
point in the same direction.

\section*{Acknowledgements}

This work was supported in part by DOE grant~DE-FG02-91ER40671, and by
a DOE Outstanding Junior Investigator award.  I thank I.~Klebanov,
H.~Verlinde, and E.~Witten for useful conversations, and the Aspen
Center for Physics for hospitality while the work was in progress.

\section*{Appendix}

We follow the notation of \cite{SchwarzIIB} except for trivial changes
in typography.  The metric signature is mostly minus, and the Clifford
algebra is $\{ \tend{\gamma}^{\flat{M}}, \tend{\gamma}^{\flat{N}} \} =
2 \eta^{\flat{MN}}$.  The notation $\tend{\gamma}$ indicates a
ten-dimensional quantity, while $\sixd{\gamma}$ indicates a
six-dimensional quantity and $\fourd{\gamma}$ indicates a
four-dimensional quantity.  We will use $M,N$ to indicate
ten-dimensional curved space indices, $\mu,\nu$ four four-dimensional
indices, and $I,J$ for six-dimensional indices.  If the
six-dimensional manifold is always assumed to have complex structure,
then $p,q$ denote three-valued holomorphic indices while
$\conj{p},\conj{q}$ denote anti-holomorphic indices.  Flat tangent
space indices are indicated by $\flat{M}$.  Symmetrization and
anti-symmetrization are carried out with ``weight one:'' for example,
$[ab] = {1 \over 2} (ab-ba)$.  For $k$-forms we use the notation
$F_{(k)} = {1 \over k!} F_{M_1 \cdots M_k} dx^{M_1} \wedge \ldots
\wedge dx^{M_k}$ where the summation over the $M_i$ is unrestricted.

A warped product geometry is a direct product of two spaces, endowed with a
metric which respects the product structure except for a conformal factor
on one of the factors which depends on the coordinates of the other.  In
our case,
  \eqn{WarpedProductMetric}{
   d\tend{s}^2 = e^{2A(y)} \fourd{g}_{\mu\nu} dx^\mu dx^\nu -
    e^{2B(y)} \sixd{g}_{IJ} dy^I dy^J \,,
  }
 where the minus sign allows us to have a six-dimensional metric with
positive signature, and for the sake of generality we have not
required $B = -A$.  (This $B$ is not to be confused with the complex
coupling in the main text.)  Hodge duals are defined as
  \eqn{HodgeDef}{
   (\tend{*} \omega)_{\flat{P}_1 \cdots \flat{P}_k} = 
    {1 \over (10-k)!} 
    \tend\epsilon_{\flat{P}_1 \cdots \flat{P}_k}{}^{\flat{P}_{k+1} 
     \cdots \flat{P}_{10}} \omega_{\flat{P}_{k+1} \cdots \flat{P}_{10}}
  }
 and similarly for the four- and six-dimensional Hodge duals $\fourd{*}$
and $\sixd{*}$.  We take the convention $\tend\epsilon^{\flat{01} \cdots
\flat{9}} = 1$ in order to agree with \cite{SchwarzIIB}; also
$\fourd\epsilon^\flat{0123} = \sixd\epsilon^\flat{456789} = 1$.  Finally,
  \eqn{volForms}{\eqalign{
   &\vol_4 = \sqrt{|\det \fourd{g}_{\mu\nu}|} dx^0 \wedge dx^1 \wedge 
    dx^2 \wedge dx^3  \qquad
   \vol_6 = \sqrt{|\det \sixd{g}_{IJ}|} dy^4 \wedge \cdots \wedge dy^9  \cr
   &\vol_{10} = e^{4A + 6B} \vol_4 \wedge \vol_6 \,.
  }}

The obvious choice of 10-bein for \WarpedProductMetric\ is
$\tend\e^\flat\mu{}_\nu = e^A \fourd\e^\flat\mu{}_\nu$ and
$\tend\e^\flat{I}{}_J = e^B \sixd\e^\flat{I}{}_J$.  The non-vanishing
components of the Christoffel connection and the spin connection are
  \eqn{ChrisSpin}{\eqalign{
   &\tend\Gamma^\nu_{\mu\lambda} = \fourd\Gamma^\nu_{\mu\lambda} \qquad
    \tend\Gamma^\nu_{I\lambda} = \delta^\nu_\lambda \partial_I A \qquad
    \tend\Gamma^J_{\mu\lambda} = -e^{2A-2B} \fourd{g}_{\mu\lambda} 
    \sixd{g}^{JK} \partial_K A  \cr
   &\tend\Gamma^J_{IK} = \sixd\Gamma^J_{IK} + 
     \delta^J_I \partial_K B + \delta^J_K \partial_I B -
     \sixd{g}_{IK} \sixd{g}^{JL} \partial_L B  \cr
   &\tend\omega_{\mu\flat{\nu\lambda}} = 
      \fourd\omega_{\mu\flat{\nu\lambda}} \qquad
    \tend\omega_{\mu\flat{\nu K}} = 
     -\tend\omega_{\mu\flat{K\nu}} = e^{A-B}
     \fourd\e_{\mu\flat\nu} \sixd\e^L{}_\flat{K} \partial_L A  \cr
   &\tend\omega_{I\flat{JK}} = -\sixd\omega_{I\flat{JK}} -
     \sixd\e_{I\flat{J}} \sixd\e^L{}_\flat{K} \partial_L B + 
     \sixd\e_{I\flat{K}} \sixd\e^L{}_\flat{J} \partial_L B \,.
  }}
 The signs in last line looks peculiar, but they are only the result
of the fact that ten-dimensional flat indices (which appear on the
left-hand side) are lowered with $\tend\eta_\flat{MN}$, while
six-dimensional flat indices (which appear on the right-hand side) are
lowered with $\delta_\flat{IJ}$, and $\tend\eta_\flat{IJ} =
-\delta_{IJ}$.  This is the penalty we pay for adopting mostly minus
signature.  It would not have been a problem if we had quoted results
for $\tend\omega_I{}^\flat{J}{}_\flat{K}$: these components of the
ten-dimensional spin connection are precisely the same as the spin
connection for the six-dimensional metric $e^{2B} \sixd{g}_{IJ}$ with
6-bein $e^{B} \sixd\e^\flat{I}{}_J$.  

The ten-dimensional gamma matrices can be expressed as
  \eqn{GammaSplit}{\eqalign{
   &\tend{\gamma}^\flat\mu = \fourd{\gamma}^\flat\mu \otimes {\bf 1} \qquad
    \tend{\gamma}^\flat{I} = 
      \fourd{\gamma}_5 \otimes \sixd{\gamma}^\flat{I}  \cr
   &\fourd{\gamma}_5 = \fourd{\gamma}_{\flat{0}} \fourd{\gamma}_{\flat{1}}
     \fourd{\gamma}_{\flat{2}} \fourd{\gamma}_{\flat{3}} \qquad
    \sixd{\gamma}_7 = \sixd{\gamma}^{\flat{4}} \sixd{\gamma}^{\flat{5}}
     \cdots \sixd{\gamma}^{\flat{9}}  \cr
   &\tend{\gamma}_{11} = \tend{\gamma}^{\flat{0}} 
     \tend{\gamma}^{\flat{1}} \cdots \tend{\gamma}^{\flat{9}} = 
     \fourd\gamma_5 \otimes \sixd\gamma_7 \,.
  }}
 By convention, the matrices $\sixd\gamma^\flat{J}$ are the same
whether one is thinking of $\flat{J}$ as a true six-dimensional index
or as a ten-dimensional index restricted to the internal manifold.
This means that $\sixd\gamma_\flat{J}$ is ambiguous in sign due to the
choice of mostly minus signature the ten-dimensional flat metric and
positive signature for the six-dimensional flat metric.  Let us choose
$\sixd\gamma_\flat{J} = \delta_\flat{JK} \sixd\gamma^\flat{K}$.
Finally, if $\omega_{(p)}$ is a $p$-form, then we define
  \eqn{OmegaSlashed}{
   \slashed\omega_{(p)} = {1 \over p!} \omega_{M_1 \cdots M_p} 
    \tend\gamma^{M_1 \cdots M_p} \,.
  }

\bibliography{Gflux}
\bibliographystyle{ssg}

\end{document}

%% file: lmacros.tex
\input umacros
\def\lbldef#1#2{\expandafter\gdef\csname #1\endcsname {#2}}
\def\eqn#1#2{\lbldef{#1}{(\ref{#1})}%
\begin{equation} #2 \label{#1} \end{equation}}
\def\eqalign#1{\vcenter{\openup1\jot
    \halign{\strut\span\TL & \span\TR\cr #1 \cr
   }}}


%% file: umacros.tex
\def\TL{\hfil$\displaystyle{##}$}
\def\TR{$\displaystyle{{}##}$\hfil}

\def\TT{\hbox{##}}
\def\seqalign#1#2{\vcenter{\openup1\jot
  \halign{\strut #1\cr #2 \cr}}}

\def\comment#1{}
\def\fixit#1{}

\def\mop#1{\mathop{\rm #1}\nolimits}

\def\vol{\mop{vol}}

\def\tr{\mop{tr}}

\def\overleftrightarrow#1{\vbox{\ialign{##\crcr
     $\leftrightarrow$\crcr\noalign{\kern-0pt\nointerlineskip}
     $\hfil\displaystyle{#1}\hfil$\crcr}}}

\def\lsim{\mathrel{\mathstrut\smash{\ooalign{\raise2.5pt\hbox{$<$}\cr\lower2.5pt\hbox{$\sim$}}}}}
\def\gsim{\mathrel{\mathstrut\smash{\ooalign{\raise2.5pt\hbox{$>$}\cr\lower2.5pt\hbox{$\sim$}}}}}

\def\slashed#1{{\ooalign{\hfil\hfil/\hfil\cr $#1$}}}

\def\sqr#1#2{{\vcenter{\vbox{\hrule height.#2pt
         \hbox{\vrule width.#2pt height#1pt \kern#1pt
            \vrule width.#2pt}
         \hrule height.#2pt}}}}
\def\square{\mathop{\mathchoice\sqr56\sqr56\sqr{3.75}4\sqr34\,}\nolimits}

\def\href#1#2{#2}  
